\documentclass[preprint,3p,onecolumn,final]{elsarticle}

\usepackage{amsmath,amssymb}
\usepackage{lineno,hyperref}

\journal{Physics Letters B}

\def\Nc{\mathrm{N_c}}
\def\CF{\mathrm{C_F}}
\def\CA{\mathrm{C_A}}
\renewcommand{\d}{\mathrm{d}}
\def\inn{\mathrm{in}}
\def\out{\mathrm{out}}
\newcommand{\as}{\alpha_s}

\begin{document}

\begin{frontmatter}

\title{\textbf{Dijet azimuthal decorrelation in \texorpdfstring{${\boldsymbol{e^+e^-}}$}{e+e-} annihilation}}

\author[Batna]{Hana Benslama}
\ead{hana.benslama@univ-batna.dz}

\author[Batna]{Yazid Delenda\texorpdfstring{\corref{cor}}}
\ead{yazid.delenda@univ-batna.dz}
\cortext[cor]{Corresponding author.}

\author[Relizane,Chlef]{Kamel Khelifa-Kerfa}
\ead{kamel.khelifakerfa@univ-relizane.dz}

\address[Batna]{Laboratoire de Physique des Rayonnements et de leurs Interactions avec la Mati\`{e}re,\\
D\'{e}partement de Physique, Facult\'{e} des Sciences de la Mati\`{e}re,\\
Universit\'{e} de Batna-1, Batna 05000, Algeria}

\address[Relizane]{D\'{e}partement de Physique, Facult\'{e} des Sciences et Technologies\\
Universit\'{e} Ahmed Zabana de Relizane, Relizane 48000, Algeria}

\address[Chlef]{Laboratoire de Math\'{e}matique et Applications,\\
Universit\'{e} Hassiba Benbouali de Chlef, Chlef 02180, Algeria}

\begin{abstract}
We examine non-global and clustering logarithms in the distribution of the azimuthal decorrelation between two jets in $e^+e^-\to$ dijet events, where the jets are defined with $E$-scheme recombination in the generalized $k_t$ algorithm. We calculate at one loop and to all orders the leading global single logarithms in the distribution of the said observable. We also compute at fixed order up to four loops at finite $\Nc$  the non-global and clustering logarithms, and numerically resum them to all orders in the large-$\Nc$ approximation. We compare our results at $\mathcal{O}(\alpha_s)$ and $\mathcal{O}(\alpha_s^2)$ with those of the \texttt{EVENT2} fixed-order Monte Carlo program and find agreement of the leading singular behavior of the azimuthal decorrelation distribution. We find that the impact of non-global logarithms on the resummed distribution in the anti-$k_t$ algorithm is substantial, while it is significantly smaller in the $k_t$ algorithm. Furthermore, the combined clustering and non-global logarithms in the $k_t$ algorithm have an even smaller effect on the distribution. Finally, we use the program \texttt{Gnole} to calculate the resummed distribution at NLL accuracy, thus achieving state-of-the-art accuracy for the resummation of this quantity.
\end{abstract}

\begin{keyword}
QCD \sep Jets \sep Resummation
\end{keyword}

\end{frontmatter}

\section{Introduction}

The production of jets in $e^+e^-$ collisions is a simple and clean environment, yet rich of physics, to test QCD and the Standard Model. It will be used in future colliders such as the ILC and FCC-ee in order to make precise measurements of QCD-related quantities, which together with detailed theoretical calculations will pave the way towards potential discovery of new-physics phenomena.

At lowest order two correlated jets are produced back-to-back with a relative azimuthal angle equal to $\pi$. At higher orders the jets manifest a decorrelation of azimuthal angle $\delta\phi$ which is enhanced near the back-to-back limit. The quantity $\delta\phi$, being sensitive to soft/collinear QCD effects, is of great interest in the phenomenology of perturbative and non-perturbative QCD dynamics. For instance it has been used to study unintegrated parton distribution functions in deep-inelastic $e-p$ scattering (DIS) \cite{Hautmann:2008vd} and small-$x$ BFKL effects \cite{H1:2003mpu}, as well as measurements of the QCD coupling at various scales \cite{ATLAS:2018sjf}. Many studies have been devoted to the distribution of $\delta\phi$ in various processes, such as  dijet production in $p-p$ \cite{ATLAS:2011kzm,Sirunyan:2017jnl,Abazov:2004hm} (and even $p-pb$ \cite{ATLAS:2019jgo}) collisions and DIS \cite{H1:2003mpu,Banfi:2008qs}. Experimentally, boson-jet (in $pp$ collisions) \cite{CMS:2013lua} and lepton-jet or photon-jet (in DIS) \cite{H1:2021wkz,Liu:2020dct,ZEUS:2017egz} decorrelations have been measured.

Near the back-to-back limit, the distribution of the azimuthal decorrelation is characterized by large logarithms preventing the convergence of the perturbative series, and thus need to be resummed to all orders. Depending on the nature of the algorithm being used to define the jets, the leading logarithms in this distribution can be double or single logarithms. For instance, in $p_t$-weighted recombination scheme of the $k_t$ \cite{Catani:1993hr,Ellis:1993tq,Cacciari:2011ma}, anti-$k_t$ \cite{Cacciari:2008gp} and Cambridge/Aachen algorithms \cite{Dokshitzer:1997in,Wobisch:1998wt}, the leading logarithms are double, $\alpha_s^n L^{2n}$, while in $E$-scheme recombination they are single, $\alpha_s^n L^n$, with $L=\ln(1/\delta\phi)$. In the former scheme, the jets recoil against emissions everywhere in the phase space, and in particular soft \emph{and} collinear emissions to these jets, which leads to the double logarithms in $\delta\phi$. On the other hand, in the latter ($E$-scheme), the jets recoil only against emissions that do not get clustered to them, and hence only away-from-jets emissions contribute to $\delta\phi$, resulting in leading soft wide-angle single logarithmic contributions.

In addition to this, the classification of the $\delta\phi$ observable, in $E$-scheme definition, falls in the ``non-global'' category, and as a consequence its distribution receives contributions from single non-global (NGLs) \cite{Dasgupta:2001sh,Dasgupta:2002bw} and/or clustering (CLs) \cite{Banfi:2005gj,Delenda:2006nf} logarithms. The resummation of these logarithms is not straightforward, and is usually performed numerically via Monte Carlo (MC) programs in the planar (large-$\Nc$) limit.

In this letter, we are interested in the calculation of NGLs and/or CLs for the $\delta\phi$ distribution both in the $k_t$ and anti-$k_t$ algorithms. We compute the coefficients of these logarithms at finite $\Nc$ as a function of the jet radius $R$ up to $\mathcal{O}(\alpha_s^3)$, and at $\mathcal{O}(\alpha_s^4)$ in the anti-$k_t$ algorithm at small $R$.  We use the fixed-order MC program \texttt{EVENT2} \cite{Catani:1996vz,Catani:1996jh} in order to compare the leading singular behavior of the $\delta\phi$ distribution with our results at $\mathcal{O}(\alpha_s)$ and $\mathcal{O}(\alpha_s^2)$. We also compute the resummed NGLs and CLs at all orders in the large-$\Nc$ limit using the MC code of refs. \cite{Dasgupta:2001sh,Dasgupta:2002bw} as well as the recently-published program \texttt{Gnole} \cite{Banfi:2021owj,Banfi:2021xzn} (in the anti-$k_t$ algorithm). The latter program is also used to compute the resummed differential $\delta\phi$ distribution at next-to-leading logarithmic (NLL) accuracy, in which we additionally control all the sub-leading logarithms $\alpha_s^{n+1} L^{n}$ in the exponent of the resummation, and quantify the corresponding scale uncertainties.

This letter is organized as follows. In the next section we compute at $\mathcal{O}(\alpha_s)$ the leading-order distribution focusing on the logarithmic contribution, and compare with fixed-order MC programs at this order. In section 3, we present the calculation of NGLs and CLs at $\mathcal{O}(\alpha_s^2)$ and show plots of the coefficients of these logarithms as a function of the jet radius and comment on the relative size of these coefficients. We also compare at this order the calculated $\delta\phi$ distribution with the output of the program \texttt{EVENT2}, thus confirming our results. In section 4 we extend the calculation to $\mathcal{O}(\alpha_s^3)$ and (in the anti-$k_t$ algorithm and at small $R$) $\mathcal{O}(\alpha_s^4)$, and point out the significantly different color structure of NGLs in $k_t$ clustering. In section 5 we present the all-orders resummation of the NGLs and/or CLs in the large-$\Nc$ limit up to LL accuracy for the $k_t$ algorithm, and NLL accuracy in the anti-$k_t$ clustering. Finally we draw our conclusions in section 6.

\section{One-loop calculation and the global form factor}
\label{sec:One-loop}

In this letter we consider the process of dijet production in $e^+e^-$ annihilation at center-of-mass energy $\sqrt{s}$. The jets are reconstructed with the $k_t$ \cite{Cacciari:2011ma} or anti-$k_t$ \cite{Cacciari:2008gp} algorithms, suited for $e^+e^-$ annihilation, with merging and stopping distances $d_{ij}$  and $d_i$ defined by
\begin{equation}\label{eq:dis}
d_{ij}= \min\left(E_i^{2p},E_j^{2p}\right)\frac{1-\cos\theta_{ij}}{1-\cos R}\,,\qquad
d_{i}= E_i^{2p}\,,
\end{equation}
where $p=+1$ for the $k_t$ algorithm and $p=-1$ for the anti-$k_t$ algorithm. Here $R$ is the jet radius, $E_i$ is the energy of the $i^\mathrm{th}$ parton in the final state, and $\theta_{ij}$ is the opening angle between partons $i$ and $j$. The algorithm sequentially merges objects $i$ and $j$ whenever $d_{ij}$ is the smallest of all merging and stopping distances, and if an object $i$ has its stopping distance as the smallest then it gets admitted to the list of final inclusive jets. The algorithm keeps recursing until all partons are clustered into jets. In this letter, we assume the jet kinematics to be defined with $E$-scheme recombination, such that the 4-momentum of a merged object simply equals the vectorial sum of the momenta of its constituents.

At the Born level, the two jets are produced back-to-back, and their relative azimuthal angle (with respect to the beam axis) is exactly $\pi$. The observable we are interested in is the deviation from $\pi$ of this relative azimuthal angle, $\delta\phi$, when soft gluons are emitted at higher orders. It is straightforward to obtain the following expression for $\delta\phi$ in terms of the transverse momenta of the emitted gluons $\kappa_{ti}$ and their azimuthal angles $\varphi_i$, with respect to the \emph{beam} axis
\begin{equation}\label{eq:def}
\delta\phi=\left|\sum_{i\notin\mathrm{jets}}\frac{\kappa_{ti}}{p_t}\sin\varphi_i\right|,
\end{equation}
where $p_{t}$ is the jet transverse momentum. The (algebraic) sum is over all emitted gluons that are not clustered to any of the two measured (leading) jets. This definition is valid only at single leading logarithmic (LL) accuracy, and we shall give the proper definition, valid at NLL accuracy, in section 5. Furthermore, the expression of $\delta\phi$ in eq. \eqref{eq:def} only applies in $E$-scheme recombination. Alternative jet recombination schemes exist for which the jet kinematics take a different form, e.g. the $p_t$-weighted scheme, and the resummation takes an entirely different structure \cite{Banfi:2008qs}.

At one loop the cumulative cross-section for events with azimuthal decorrelation $\delta\varphi$ less than some $\Delta$, normalized to the Born cross-section, reads
\begin{equation}
\Sigma_1(\Delta)=-2\,\CF\int\frac{\alpha_s(k_t)}{\pi}\,\frac{\d k_t}{k_t}\,\frac{\d\cos\theta}{\sin^2\theta}\,\frac{\d\phi}{2\pi}\,\omega_{q\bar{q}}^k\, \Theta_{\out}(k)\,\Theta\left(\kappa_t|\sin\varphi|/p_t-\Delta\right),
\end{equation}
where $\theta$, $\phi$ and $k_t$ are the polar angle, azimuthal angle and transverse momentum of the emitted soft gluon $k$, with respect to the \emph{jet} (thrust) axis (the back-to-back outgoing jets are aligned along the $z$ axis), $\CF$ is the color factor associated with the emission of the gluon off the hard $q\bar{q}$ dipole, and $\as$ is the strong coupling with argument $k_t$. The invariant antenna function $\omega_{q\bar{q}}^k$ is given by
\begin{equation}
\omega_{q\bar{q}}^k=\frac{k_t^2}{2}\,\frac{p_q\cdot p_{\bar{q}}}{(p_q\cdot k)(p_{\bar{q}}\cdot k)}=1\,,
\end{equation}
with $p_i$ denoting the momentum of particle $i$
\begin{subequations}
\begin{align}
p_q&=\frac{\sqrt{s}}{2}(1,0,0,1)\,,\\
p_{\bar{q}}&=\frac{\sqrt{s}}{2}(1,0,0,-1)\,,\\
k&=E\left(1,\cos\phi \sin\theta, \sin\phi \sin \theta, \cos\theta\right),
\end{align}
\end{subequations}
where $k_t = E \sin \theta$. The constraint $\Theta_{\out}(k)$ restricts the emitted gluon to be outside the jets and forbids it from being clustered to any of them in order to induce a non-zero azimuthal decorrelation. It depends on the jet radius $R$ and is given, in the generalized algorithm, by
\begin{equation}
\Theta_{\out}(k)=\Theta\left(\cos R-|\cos\theta|\right).
\end{equation}

Since the soft emission is restricted to be outside both jets then there are no collinear logarithms associated with this observable. This means that the leading logarithms are single, which allows us at LL accuracy to simply change $\Theta(\kappa_t|\sin\varphi|/p_t-\Delta)\to\Theta(2\,k_t/\sqrt{s}-\Delta)$, since any factor multiplying $k_t$ will only induce sub-leading logarithms. \footnote{Notice that $\kappa_t$ and $\varphi$ are different from  $k_t$ and $\phi$.} We can then perform the integration over $k_t$ using the one-loop running of the coupling (which formally enters the distribution at higher orders), and write the result in terms of the evolution parameter $t$ defined by
\begin{equation}\label{eq:EvolParam}
t(\Delta)\equiv\int^{\sqrt{s}/2}\frac{\alpha_s(k_t)}{\pi}\,\frac{\d k_t}{k_t}\,\Theta\left(2\,k_t/\sqrt{s}-\Delta\right)= -\frac{1}{2\pi\beta_0}\ln(1-2\lambda)\,,
\end{equation}
where $\lambda=\alpha_s(\sqrt{s}/2)\beta_0\ln(1/\Delta)$ and $\beta_0$ is the one-loop coefficient of the QCD beta function. The angular integration is straightforward and we obtain
\begin{equation}\label{eq:1loop}
\Sigma_1(\Delta)=-2\,\CF\,t(\Delta)\int_{-1}^1\frac{\d c}{1-c^2}\,\Theta\left(\cos R-|c|\right)=-2\,\CF\,t(\Delta)\ln\frac{1+\cos R}{1-\cos R}\,,
\end{equation}
with $c$ standing for $\cos\theta$. Note that since only emissions in the inter-jet (gap) region are integrated over, this result may be cast in terms of the rapidity-gap width $\Delta\eta$
\begin{equation}\label{eq:gap}
\Delta\eta \equiv \ln\frac{1+\cos R}{1-\cos R}\,.
\end{equation}
The all-orders resummed global form factor is simply the exponential of the one-loop distribution. That is
\begin{equation}\label{eq:ResummedGlobalFormFactor}
\Sigma_{\mathrm{global}}(\Delta)=\exp\left[-2\,\CF\,t(\Delta)\,\Delta\eta\right].
\end{equation}
An identical expression was also arrived at in ref. \cite{Khelifa-Kerfa:2011quw} for jet shapes in $e^+e^-$ annihilation.

\begin{center}
\begin{figure}[ht]
\centering
\includegraphics[width=0.45\textwidth]{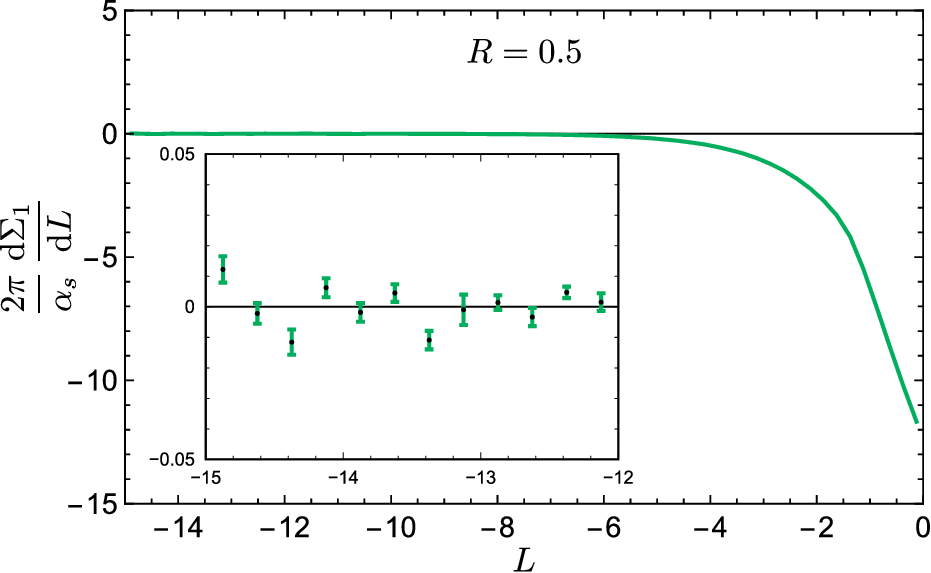}
\caption{\label{fig:00}The difference between the leading-order \texttt{EVENT2} differential distribution $2\pi/\alpha_s\,\d\Sigma_1/\d L$ and the resummed distribution expanded at $\mathcal{O}(\alpha_s)$. The singular behavior of the MC distribution is exactly canceled by the expanded result.}
\end{figure}
\end{center}

The leading-order result can be verified by comparing it with the output of the MC program \texttt{EVENT2} at $\mathcal{O}(\alpha_s)$ \cite{Catani:1996vz,Catani:1996jh}. Specifically we compare the differential distribution
\begin{equation}
\frac{2\pi}{\alpha_s}\,\frac{\d\Sigma_1}{\d L}=4\,\CF\,\ln\frac{1+\cos R}{1-\cos R}\,,
\end{equation}
where $L=\ln\Delta$, with the same MC distribution, for the chosen value of $R=0.5$. We show in figure \ref{fig:00} a plot of the difference between the MC distribution and the expansion of the resummation at $\mathcal{O}(\alpha_s)$, where, as expected, this difference tends to zero in the logarithmically-enhanced region.

\section{Two-loops calculation: NGLs and CLs}

When employing the $k_t$ or anti-$k_t$ clustering algorithms with $E$-scheme recombination, the resummation of the azimuthal decorrelation distribution requires the treatment of NGLs and/or CLs. The corresponding cumulative distribution at $\mathcal{O}(\alpha_s^2)$ can then be split into three contributions
\begin{equation}\label{eq:2lpss}
\Sigma_2(\Delta) = \frac{1}{2!}\left[\Sigma_1(\Delta)\right]^2+\Sigma_2^{\mathrm{NG}}(\Delta)+ \Sigma_2^{\mathrm{CL}}(\Delta)\,,
\end{equation}
with $\Sigma_2^{\mathrm{CL}}(\Delta)=0$ for anti-$k_t$ clustering. Let us first discuss the NGLs contribution $\Sigma_2^{\mathrm{NG}}(\Delta)$ in both algorithms, and then compute the CLs contribution $\Sigma_2^{\mathrm{CL}}(\Delta)$ for $k_t$ clustering.

\subsection{Calculation of NGLs}

The origin of NGLs  at two loops is the emission of a soft gluon $k_1$ inside any of the two outgoing jets which itself emits a softer gluon $k_2$ outside the jets without being clustered back to them. While this configuration results in a non-zero $\delta\phi$, its virtual correction (specifically when $k_2$ is virtual) gives $\delta\phi=0$, and thus we have a real-virtual mis-cancellation of the soft singularities. We express the contribution of the uncanceled virtual correction to the integrated azimuthal decorrelation distribution as follows
\begin{subequations}
\begin{align}
\Sigma_2^{\mathrm{NG}}(\Delta)&=\mathcal{S}_2(R)\,\frac{t^2}{2!}\,,\\
\mathcal{S}_2(R)&=-2\,\CF\,\CA\int\frac{\d c_1}{1-c_1^2}\,\frac{\d\phi_1}{2\pi}\,\frac{\d c_2}{1-c_2^2}\,\frac{\d\phi_2}{2\pi}\,\mathcal{A}_{q\bar{q}}^{12}\,\Xi^{\mathrm{NG}}_2(R)\,,\label{eq:NGLs2}
\end{align}
\end{subequations}
where $\CA$ is the color factor associated with the non-Abelian emission of gluon $k_2$ off $k_1$. The irreducible two-loops antenna function $\mathcal{A}_{q\bar{q}}^{12}$ is given by \cite{Delenda:2015tbo}
\begin{equation}\label{eq:A12}
\mathcal{A}_{q\bar{q}}^{12}=\omega_{q\bar{q}}^{1}\left(\omega_{q1}^{2}+\omega_{1\bar{q}}^{2}-\omega_{q\bar{q}}^{2}\right)=\frac{1-c_1c_2}{1-c_1c_2-s_1s_2\cos(\phi_1-\phi_2)}-1\,,
\end{equation}
with $s_i\equiv \sin\theta_i$. The function $\Xi^{\mathrm{NG}}_2$ restricts the angular phase-space of integration and is given, in the anti-$k_t$ and $k_t$ algorithms respectively, by
\begin{subequations}
\begin{align}
\Xi^{\mathrm{NG},\,\mathrm{ak}_t}_2 &= \Theta_{\inn}(k_1)\Theta_{\out}(k_2)\,,\\
\Xi^{\mathrm{NG},\,\mathrm{k}_t}_2  &= \Theta_{\inn}(k_1)\Theta_{\out}(k_2)\Theta(d_{12}-d_2)\,.\label{eq:2lpskt}
\end{align}
\end{subequations}
The step function $\Theta(d_{12}-d_2)$ forbids gluon $k_2$ from being clustered back to the jet in the $k_t$ algorithm. It is given by $\Theta(\cos R-\cos\theta_{12})$, with $\cos\theta_{12} = c_1c_2+s_1s_2\cos(\phi_1-\phi_2)$.


In the anti-$k_t$ algorithm the integration is simple and its result can be expressed in the same form as that of the rapidity-gap NGLs coefficient found in ref. \cite{Dasgupta:2002bw}
\begin{align}
\mathcal{S}^{\mathrm{ak}_t}_2(R)&=-\CF\,\CA\left[\frac{\pi^2}{6}+2\Delta\eta^2-2\Delta\eta\ln\left(e^{2\Delta \eta}-1\right)-\text{Li}_2\left(e^{-2\Delta\eta}\right)-\text{Li}_2\left(1-e^{2\Delta\eta}\right)\right]\notag\\
&=-\CF\,\CA\left[\frac{\pi^2}{3}-\frac{R^4}{8}-\frac{R^6}{24}-\frac{29\,R^8}{2560}+\mathcal{O}\left(R^{10}\right)\right],
\end{align}
where $\Delta\eta(R)$ is defined in eq. \eqref{eq:gap}. The above formula shows that in the limit $R\to0$ the two-loops NGLs coefficient does not vanish, but rather reaches its maximum value. This feature was observed in ref. \cite{Dasgupta:2002bw} and was ascribed to the fact that NGLs originate from the \emph{edge} of jets, since this is the phase-space region where gluons $k_1$ and $k_2$ are collinear and thus the amplitude squared \eqref{eq:A12} is most singular.

In the $k_t$ algorithm, and at small values of $R$ (using small angles), we can write the NGLs coefficient as
\begin{align}
\mathcal{S}^{\mathrm{k}_t}_2(R\sim0)&=-8\,\CF\,\CA\,\int_0^{1}\d\theta_1\int_1^{\infty}\d\theta_2\int_0^{2\pi}\frac{\d\phi}{2\pi}\,\frac{1}{\left(\theta_1^2+\theta_2^2\right)\sec\phi-2\,\theta_1\,\theta_2} \,\Theta\left(\theta_1^2+\theta_2^2-2\,\theta_1\,\theta_2\cos\phi-1\right)\notag\\
&=-\frac{2\pi^2}{27}\,\CF\,\CA\,,
\end{align}
where we made the following changes of variables: $\phi=\phi_1-\phi_2$ and $\theta_i\to R\,\theta_i$. Away from the small-$R$ limit one may perform the integration numerically to obtain the full-$R$ result for the two-loops coefficient of NGLs. We present the results in the following subsection together with the CLs coefficient.

\subsection{CLs with \texorpdfstring{$k_t$}{kt} clustering}


To compute the CLs we consider the Abelian primary emission of two strongly-ordered gluons directly off the hard $q\bar{q}$ dipole, whereby the harder gluon $k_1$ is inside one of the two jets and the softer $k_2$ is outside both of them, with the constraint $d_{12}<d_{2}$, such that gluon $k_2$ gets clustered into the jet by gluon $k_1$, which leads to $\delta\phi=0$. However, when $k_1$ is virtual then gluon $k_2$ remains in the gap causing the hard jets to decorrelate. In this case we obtain a large single logarithmic contribution to the $\delta\phi$ distribution given by
\begin{subequations}
\begin{align}
&\Sigma_2^{\mathrm{CL}}(\Delta)=\mathcal{C}^{\mathrm{k}_t}_2(R)\,\frac{t^2}{2!}\,,\\
&\mathcal{C}^{\mathrm{k}_t}_2(R)=4\,\mathrm{C_F^2}\int \frac{\d c_1}{1-c_1^2}\,\frac{\d\phi_1}{2\pi}\,\frac{\d c_2}{1-c_2^2}\,\frac{\d\phi_2}{2\pi}\,w_{q\bar{q}}^{1}\,w_{q\bar{q}}^{2}\,
\Theta(|c_1|-\cos R)\,\Theta(\cos R-|c_2|)\,\Theta(\cos\theta_{12}-\cos R)\,.
\end{align}
\end{subequations}
Note again that $\mathcal{C}^{\mathrm{ak}_t}_2(R)=0$, as there are no CLs for anti-$k_t$ clustering.

\begin{center}
\begin{figure}[ht]
\centering
\includegraphics[width=0.45\textwidth]{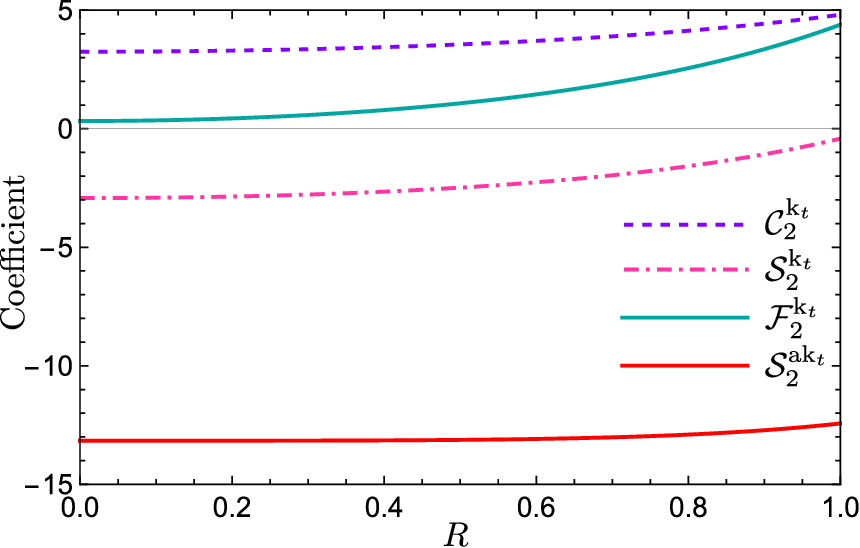}
\caption{\label{fig:C2}CLs and NGLs coefficients at two loops with $k_t$ and anti-$k_t$ clustering.}
\end{figure}
\end{center}

First, let us consider the small-$R$ limit of this integral. In this case we may write
\begin{equation}
\mathcal{C}^{\mathrm{k}_t}_2(R\sim0)=8\,\mathrm{C_F^2}\int_0^1 \frac{\d \theta_1}{\theta_1}\,\int_1^{\infty}\frac{\d \theta_2}{\theta_2}\int_0^{2\pi} \frac{\d\phi}{2\pi}\,\Theta\left(-\theta_1^2-\theta_2^2+2\,\theta_1\,\theta_2\cos\phi+1\right)= \frac{5\pi^2}{27}\,\mathrm{C_F^2}\,.
\end{equation}
Away from this small-$R$ limit we perform the integration numerically. We show in figure \ref{fig:C2} a plot of the coefficients of NGLs and CLs at $\mathcal{O}(\alpha_s^2)$ as a function of the jet radius $R$. Also shown is the combined coefficient of CLs and NGLs in the $k_t$ algorithm; $\mathcal{F}^{\mathrm{k}_t}_2=\mathcal{S}^{\mathrm{k}_t}_2+\mathcal{C}^{\mathrm{k}_t}_2$.

Notice that the NGLs coefficient in $k_t$ clustering is significantly smaller than that in anti-$k_t$. Physically, this can be explained by the fact that the $k_t$ clustering procedure reduces the angular phase space available for collinear enhancement of the two gluons near the boundaries of the jets. This reduction of NGLs was reported even at NLL accuracy \cite{Khelifa-Kerfa:2011quw}. The CLs coefficient is positive and quite small, with the advantage that it cancels the NGLs coefficient, particularly at small values of $R$. Note that the overall coefficient $\mathcal{F}_2$ is less than 1 for values of $R$ smaller than about $0.5$. For small $R$ values, the anti-$k_t$ NGLs coefficient computed here is identical to that found for the single hemisphere mass \cite{Banfi:2010pa} and jet mass with a jet veto \cite{Khelifa-Kerfa:2011quw}.

\subsection{Comparison to EVENT2}

\begin{center}
\begin{figure}[ht]
\centering
\includegraphics[height=3.6cm]{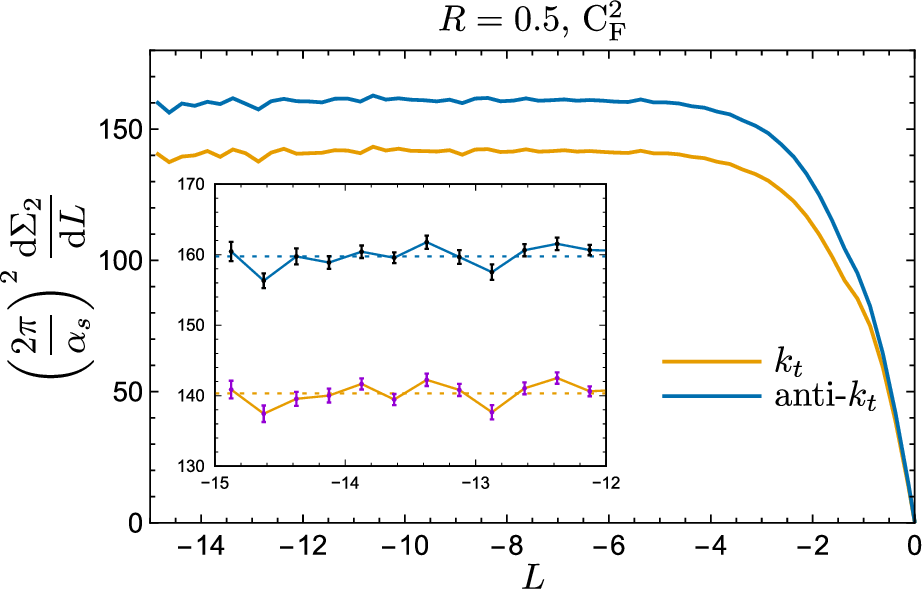}
\includegraphics[height=3.6cm]{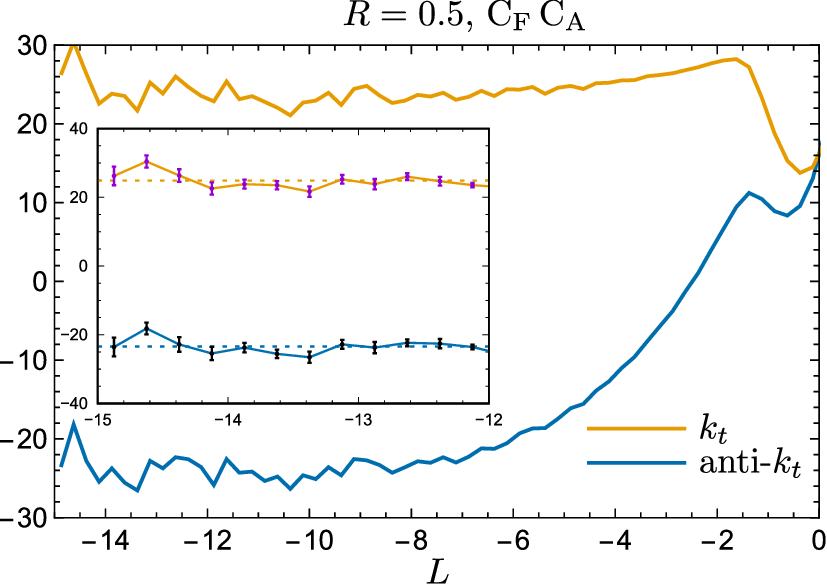}
\includegraphics[height=3.6cm]{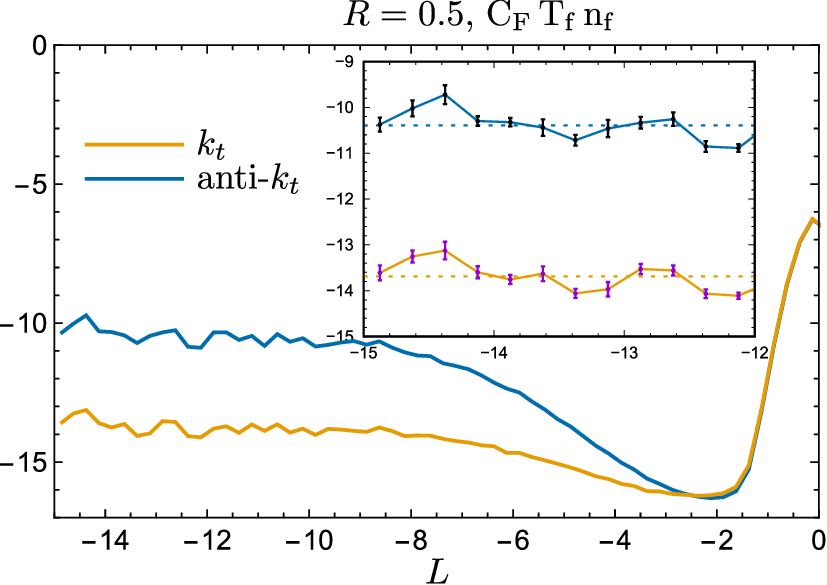}
\caption{\label{fig:231ps}The difference between the NLO \texttt{EVENT2} differential distribution $(2\pi/\alpha_s)^2\,\d\Sigma_2/\d L$ and the resummed distribution expanded at $\mathcal{O}(\alpha_s^2)$. The leading logarithmic behavior of the MC distribution $\mathcal{O}(\alpha_s^2 L)$ is canceled, leaving a constant behavior at large values of $L$.}
\end{figure}
\end{center}

We compare our two-loops results with the exact MC distribution at $\mathcal{O}(\alpha_s^2)$ obtained with the \texttt{EVENT2} program. The latter splits the distribution at NLO into three color contributions; $\mathcal{O}(\mathrm{C_F^2})$, $\mathcal{O}(\CF\,\CA)$ and $\mathcal{O}(\CF\,\mathrm{T_f}\,\mathrm{n_f})$, with $\mathrm{T_f}=1/2$ being the normalization constant of the generators of the SU(3) group in the fundamental representation and $\mathrm{n_f}=5$ is the number of active quark flavors. The expansion of the resummed distribution at this order, including running coupling effects, is written, after differentiating with respect to $L$, as
\begin{equation}\label{eq:Expd}
\left(\frac{2\pi}{\alpha_s}\right)^2\frac{\d\Sigma_2}{\d L}=\left[\mathrm{C_F^2}\,4\left(4\,\Delta\eta^2+\mathcal{C}_2\right)
-\CF\,\CA\,\frac{4}{3}\left(3\,\mathcal{S}_2+11\Delta\eta\right)+\CF\,\mathrm{T_f}\,\mathrm{n_f}\,\frac{16}{3}\,\Delta\eta\right]L+\mathcal{O}(1)\,.
\end{equation}
At this order this differential distribution is linear in $L$, but does not capture the $\mathcal{O}(1)$ constant, which, in the cumulative integrated distribution, is an NLL $\mathcal{O}(\alpha_s^2 \,L)$ term. We plot the difference between the MC distribution and the expansion \eqref{eq:Expd} for each color contribution separately. All curves should tend towards a constant when $L$ grows large and negative. The results are shown in figure \ref{fig:231ps}. Here the $\mathcal{O}(\mathrm{C_F^2})$ part contains the CLs coefficient in $k_t$ clustering, while the $\mathcal{O}(\CF\,\CA)$ term contains the NGLs coefficient both in $k_t$ and anti-$k_t$ algorithms. The $\mathcal{O}(\mathrm{C_F\,T_f\,n_f})$ contribution is algorithm independent at LL accuracy, but the NLL $\mathcal{O}(1)$ constant does depend on the jet algorithm.

\section{NGLs and CLs at three loops}

\subsection{NGLs at three loops}

At $\mathcal{O}(\alpha_s^3)$, the cumulative distribution can be written as follows
\begin{equation}\label{eq:Sig3}
\Sigma_3(\Delta)=\frac{1}{3!}\left[\Sigma_1(\Delta)\right]^3
+\Sigma_1(\Delta)\times\Sigma_2^{\mathrm{NG}}(\Delta)+\Sigma_1(\Delta)\times\Sigma_2^{\mathrm{CL}}(\Delta)+\Sigma_3^{\mathrm{NG}}(\Delta)+\Sigma_3^{\mathrm{CL}}(\Delta)\,.
\end{equation}
Focusing first on the \emph{pure} NGLs contribution $\Sigma_3^{\mathrm{NG}}$ (i.e., excluding the cross-talk between the one-loop global and two-loops non-global logarithms) in the anti-$k_t$ algorithm, we may write it in the form
\begin{subequations}
\begin{align}
&\Sigma_3^{\mathrm{NG,ak}_t}(\Delta)=\mathcal{S}_3^{\mathrm{ak}_t}(R)\,\frac{t^3}{3!}\,,\label{eq:Sig3akt}\\
&\mathcal{S}_3^{\mathrm{ak}_t}(R)=2\,\CF\,\mathrm{C_A^2}\int\left(\prod_{i=1}^3\frac{\d c_i}{1-c_i^2}\,\frac{\d\phi_i}{2\pi}\right)\left[\mathcal{A}_{q\bar{q}}^{12}\,\bar{\mathcal{A}}_{q\bar{q}}^{13}\,\Theta_\inn(k_1) \Theta_\out(k_2)\Theta_\out(k_3)-\mathcal{B}_{q\bar{q}}^{123}\,\Theta_\inn(k_1)\Theta_\inn(k_2)\Theta_\out(k_3)\right],\label{eq:NGLs3}
\end{align}
\end{subequations}
where we define $\bar{\mathcal{A}}_{q\bar{q}}^{13}=\mathcal{A}_{q\bar{q}}^{13}/\omega_{q\bar{q}}^1$, and the 3-loops irreducible cascade antenna function is given by
\begin{equation}
\mathcal{B}_{q\bar{q}}^{123}=\omega_{q\bar{q}}^1\left[\mathcal{A}_{q1}^{23}+\mathcal{A}_{1\bar{q}}^{23}-\mathcal{A}_{q\bar{q}}^{23}\right].
\end{equation}
It is worth mentioning that, for the cascade term, there is a non-negligible contribution from the configuration in which gluon $k_1$ is in one jet emitting gluon $k_2$ in the other jet which itself emits the softest gluon $k_3$ in the gap between the two jets. At small values of $R$ the integration is quite simple and yields the result
\begin{equation}
\mathcal{S}_3^{\mathrm{ak}_t}(R\sim0)=\CF\,\mathrm{C_A^2}\,2\,\zeta_3\,.
\end{equation}
Notice that this result is twice that found for the hemisphere mass observable \cite{Khelifa-Kerfa:2015mma}. Away from the small-$R$ limit the integration can be performed numerically and we shall present the results in the next subsection.

The \emph{pure} NGLs contribution in the $k_t$ algorithm is given by an identical form to that of the anti-$k_t$ \eqref{eq:Sig3akt}
\begin{equation}
\Sigma_3^{\mathrm{NG,k}_t}(\Delta)=\mathcal{S}_3^{\mathrm{k}_t}(R)\,\frac{t^3}{3!}\,. \label{eq:Sig3kt}
\end{equation}
We perform for the first time in the literature a calculation of NGLs in the $k_t$ algorithm beyond two loops. Due to non-linearity of $k_t$ clustering we shall find a class of NGLs that have a non-standard color factor, namely $\mathrm{C_F^2}\,\CA$ at $\mathcal{O}(\alpha_s^3)$. Such terms usually (in anti-$k_t$ clustering) only arise in the cross-talk of the expansion of the primary-emission global form factor \eqref{eq:ResummedGlobalFormFactor} at $\mathcal{O}(\alpha_s)$ together with NGLs at $\mathcal{O}(\alpha_s^2)$, i.e., the term $\Sigma_1(\Delta)\times\Sigma_2^{\mathrm{NG}}(\Delta)$ in eq. \eqref{eq:Sig3}. However, in the $k_t$ algorithm, we find them as ``pure'' irreducible NGLs. That is, they are part of the term $\Sigma_3^{\mathrm{NG}}(\Delta)$ in eq. \eqref{eq:Sig3}.

\begin{center}
\begin{figure}[ht]
\centering
\includegraphics[width=0.45\textwidth]{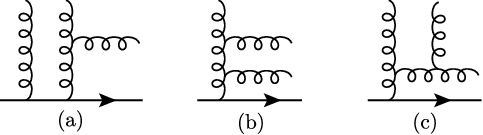}
\caption{\label{fig:3lp}The three types of emissions to consider for NGLs calculation at $\mathcal{O}(\alpha_s^3)$: (a) one primary + two correlated emissions, (b) ladder emissions, and (c) cascade emissions.}
\end{figure}
\end{center}

To proceed, we consider three types of emissions at $\mathcal{O}(\alpha_s^3)$, as shown in figure \ref{fig:3lp}: (a) one primary + two correlated emissions, (b) ladder emissions, and (c) cascade emissions. In each case we consider all possible virtual-correction Feynman diagrams as well as angular configurations of the emitted gluons that affect the clustering procedure, and look for a mis-match between the soft divergences of these emissions.

\begin{center}
\begin{figure}[ht]
\centering
\includegraphics[width=0.6\textwidth]{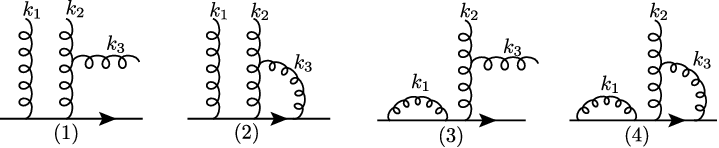}
\caption{\label{fig:32p}Feynman diagrams corresponding to the squared amplitude $4\,\mathrm{C_F^2}\,\CA\,\omega_{q\bar{q}}^1\,\mathcal{A}_{q\bar{q}}^{23}$.}
\end{figure}
\end{center}

Starting first with type (a) contributions, there are three possible permutations of the gluons. \footnote{Note that for all permutations the transverse momenta of the three gluons are strongly ordered as follows: $k_{t1} \gg k_{t2} \gg k_{t3}$.} For the permutation in which $k_3$ is emitted in correlation with $k_2$, and which has a squared amplitude $4\,\mathrm{C_F^2}\,\CA\,\omega_{q\bar{q}}^1\,\mathcal{A}_{q\bar{q}}^{23}$, we find that the angular phase space of integration that yields a logarithmic contribution is given by
\begin{align}
\Xi^{\mathrm{NG,k}_t}_{31}(R)&=\Theta_\inn(k_1)\Theta_\inn(k_2)\Theta_\out(k_3)\Theta(d_3-d_{13})\Theta(d_{23}-d_3)+\notag\\
&+\Theta_\out(k_1)\Theta_\inn(k_2)\Theta_\out(k_3)\Theta(d_{23}-d_3)+\notag\\
&-\Theta_\inn(k_1)\Theta_\out(k_2)\Theta_\out(k_3)\Theta(d_{13}-d_3)\Theta(d_{23}-d_3)\Theta(d_{2}-d_{12})\,.\label{eq:nds}
\end{align}
To see how one obtains this result let us give one example of angular configurations that result in a logarithmic contribution. There are four Feynman diagrams in this case, shown in figure \ref{fig:32p}.
Consider the situation when particles $k_3$ and $k_2$ are outside the jet regions, $d_{3j}>d_3$ and $d_{2j}>d_2$, while particle $k_1$ is inside, $d_{1j}<d_1$. In this scenario, both diagrams in which $k_1$ is virtual (i.e., diagrams (3) and (4) in figure \ref{fig:32p}) yield $\delta\phi\neq 0$, since in both diagrams particle $k_2$ is real and remains in the gap region after applying the clustering. However, these two diagrams contribute equally and with opposite signs, so they cancel each other. For the remaining two diagrams, (1) and (2), we have a mismatch when particle $k_2$ gets pulled inside the jet by the real particle $k_1$ while $k_3$ remains in the gap. This happens when $d_{12}$ is smaller than $d_2$, and both $d_{13}$ and $d_{23}$ are greater than $d_3$. While in diagram (1) we have a real unclustered gluon $k_3$ in the gap region (i.e., it forms a jet on its own) giving $\delta\phi\neq 0$, in diagram (2) the gap is empty and the hard jets are exactly back-to-back with $\delta\phi=0$. The virtual-correction diagram (2) contributes fully to the cumulative distribution while the real-emission diagram (1) cancels this contribution only up to $\delta\phi=\Delta$, leaving uncanceled virtual-correction contributions with a negative sign. This is the last term in eq. \eqref{eq:nds}.

Similarly, we obtain for the second and third gluon permutations of type (a) diagrams, with squared amplitudes $4\,\mathrm{C_F^2}\,\CA\,\omega_{q\bar{q}}^2\,\mathcal{A}_{q\bar{q}}^{13}$ and $4\,\mathrm{C_F^2}\,\CA\,\omega_{q\bar{q}}^3\,\mathcal{A}_{q\bar{q}}^{12}$, respectively, as well as type (b) (ladder-emission) contributions, with squared amplitude $2\,\CF\,\mathrm{C_A^2}\,\bar{\mathcal{A}}_{q\bar{q}}^{12} \,\mathcal{A}_{q\bar{q}}^{13}$, the same phase space function. It is given by
\begin{align}
\Xi^{\mathrm{NG,k}_t}_{32}(R)&=\Theta_\inn(k_1)\Theta_\inn(k_2)\Theta_\out(k_3)\Theta(d_{13}-d_3)\Theta(d_3-d_{23})+\notag\\
&+\Theta_\inn(k_1)\Theta_\out(k_2)\Theta_\out(k_3)\Theta(d_{13}-d_3)\Theta(d_3-d_{23})+\notag\\
&+\Theta_\inn(k_1)\Theta_\out(k_2)\Theta_\out(k_3)\Theta(d_{13}-d_3)\Theta(d_{23}-d_3)\Theta(d_{12}-d_2)\,.\label{eq:ndd}
\end{align}
Finally, for type (c) (cascade emission) contributions, corresponding to the squared amplitude $2\,\CF\,\mathrm{C_A^2}\,\mathcal{B}_{q\bar{q}}^{123}$, the phase space function reads
\begin{align}
\Xi^{\mathrm{NG,k}_t}_{33}(R)=&-\Theta_\inn(k_1)\Theta_\inn(k_2)\Theta_\out(k_3)\Theta(d_{13}-d_3)\Theta(d_{23}-d_3)-\notag\\
&-\Theta_\inn(k_1)\Theta_\out(k_2)\Theta_\out(k_3)\Theta(d_{13}-d_3)\Theta(d_{23}-d_3)\Theta(d_{2}-d_{12})\,.
\end{align}

Before performing the integration, we subtract off the part of the phase space that produces the interference term between the one-loop global primary logarithm and the two-loops NGLs, which only comes from type (a) emissions. For the first permutation of gluons, this part of phase space is identified by the second term in eq. \eqref{eq:nds}, $\Theta_\out(k_1)\Theta_\inn(k_2)\Theta_\out(k_3)\Theta(d_{23}-d_3)$, where gluon $k_1$ reproduces the one-loop global term \eqref{eq:1loop} and the other correlated gluons, $k_2$ and $k_3$, give the two-loops NGLs (eq. \eqref{eq:NGLs2} with phase space \eqref{eq:2lpskt}). However, for the other two gluon permutations of type (a) emissions, the phase space \eqref{eq:ndd} does not simply contain such interference terms. Strictly speaking, this means that NGLs do not cleanly factorize from the global form factor in the $k_t$ algorithm. Nevertheless, we can manually add and subtract the interference terms and write the total distribution in the factorizable form \eqref{eq:Sig3}.

We can then write the ``pure'' NGLs coefficient $\mathcal{S}_3^{\mathrm{k}_t}$, given in eq. \eqref{eq:Sig3kt}, in the following form
\begin{equation}
\mathcal{S}_3^{\mathrm{k}_t}(R)=\mathcal{S}_3^{(a)}(R)+\mathcal{S}_3^{(b)+(c)}(R)\,,\label{eq:NGLs3kt}
\end{equation}
where we split the result according to the color factor, such that for type (a) emissions we have
\begin{align}
&\mathcal{S}_3^{(a)}(R)=4\,\mathrm{C_F^2}\,\CA\int\left(\prod_{i=1}^3\frac{\d c_i}{1-c_i^2}\,\frac{\d\phi_i}{2\pi}\right)
\left[\omega_{q\bar{q}}^1\,\mathcal{A}_{q\bar{q}}^{23}\,\widetilde{\Xi}_{31}(R)
     +\omega_{q\bar{q}}^2\,\mathcal{A}_{q\bar{q}}^{13}\,\widetilde{\Xi}_{32}(R)
     +\omega_{q\bar{q}}^3\,\mathcal{A}_{q\bar{q}}^{12}\,\widetilde{\Xi}_{33}(R)\right],
\end{align}
with modified phase space that subtracts away the interference terms
\begin{subequations}
\begin{align}
&\widetilde{\Xi}_{31}(R)=\Xi^{\mathrm{NG,k}_t}_{31}(R)-\Theta_\out(k_1)\Theta_\inn(k_2)\Theta_\out(k_3)\Theta(d_{23}-d_3)\,,\\
&\widetilde{\Xi}_{32}(R)=\Xi^{\mathrm{NG,k}_t}_{32}(R)-\Theta_\inn(k_1)\Theta_\out(k_2)\Theta_\out(k_3)\Theta(d_{13}-d_3)\,,\\
&\widetilde{\Xi}_{33}(R)=\Xi^{\mathrm{NG,k}_t}_{32}(R)-\Theta_\inn(k_1)\Theta_\out(k_2)\Theta_\out(k_3)\Theta(d_{12}-d_2)\,,
\end{align}
\end{subequations}
and for type (b) and (c) emissions
\begin{align}
&\mathcal{S}_3^{(b)+(c)}(R)=2\,\CF\,\mathrm{C_A^2}\int\left(\prod_{i=1}^3\frac{\d c_i}{1-c_i^2}\,\frac{\d\phi_i}{2\pi}\right)\left[\mathcal{A}_{q\bar{q}}^{12}\,\bar{\mathcal{A}}_{q\bar{q}}^{13}\,\Xi^{\mathrm{NG,k}_t}_{32}(R)+\mathcal{B}_{q\bar{q}}^{123}\,\Xi^{\mathrm{NG,k}_t}_{33}(R)\right].
\end{align}

We are now in a position to perform the integrations numerically as a function of the jet radius $R$. We show the results in the next subsection, in which we also compute the clustering logarithmic contribution at $\mathcal{O}(\alpha_s^3)$.

\subsection{CLs with \texorpdfstring{$k_t$}{kt} clustering at three loops}

Following the same steps as for NGLs calculation, the phase space clustering function at $\mathcal{O}(\alpha_s^3)$ for the CLs contribution, which results from the mismatch of soft singularities between real and virtual emissions of three primary soft gluons, with a squared amplitude $8\,\mathrm{C_F^3}\,w_{q\bar{q}}^1\,w_{q\bar{q}}^2\,w_{q\bar{q}}^3$, is given by
\begin{align}
\Xi_3^{\mathrm{CL,k}_t}(R)=&-\Theta_\inn(k_1)\Theta_\inn(k_2)\Theta_\out(k_3)\Theta(d_3-d_{13})\Theta(d_3-d_{23})-\notag\\
&-\Theta_\out(k_1)\Theta_\inn(k_2)\Theta_\out(k_3)\Theta(d_3-d_{23})-\notag\\
&-\Theta_\inn(k_1)\Theta_\out(k_2)\Theta_\out(k_3)\Theta(d_3-d_{13})-\notag\\
&-\Theta_\inn(k_1)\Theta_\out(k_2)\Theta_\out(k_3)\Theta(d_{13}-d_3)\Theta(d_{23}-d_3)\Theta(d_2-d_{12})\,.
\end{align}
Note that the above phase-space clustering function is similar (but not exactly identical) to that found for the jet mass observable \cite{Delenda:2012mm}. Extracting the interference terms between the CLs at two loops and the global logarithm at one loop, i.e., the term $\Sigma_1(\Delta)\times\Sigma_2^{\mathrm{CL}}(\Delta)$ in eq. \eqref{eq:Sig3}, we reduce the above phase space function to that of the ``pure'' CLs contribution as
\begin{align}
\widetilde{\Xi}_3^{\mathrm{CL,k}_t}(R)=&-\Theta_\inn(k_1)\Theta_\inn(k_2)\Theta_\out(k_3)\Theta(d_3-d_{13})\Theta(d_3-d_{23})+\notag\\
&+\Theta_\inn(k_1)\Theta_\out(k_2)\Theta_\out(k_3)\Theta(d_2-d_{12})[1-\Theta(d_{13}-d_3)\Theta(d_{23}-d_3)]\,.
\end{align}
Then, the clustering logarithmic contribution to the cumulative cross-section is given by
\begin{subequations}
\begin{align}
&\Sigma_3^{\mathrm{CL}}(\Delta)=\mathcal{C}_3^{\mathrm{k}_t}(R)\,\frac{t^3}{3!}\,,\\
&\mathcal{C}_3^{\mathrm{k}_t}(R)=8\,\mathrm{C_F^3}\int\left(\prod_{i=1}^3\frac{\d c_i}{1-c_i^2}\,\frac{\d\phi_i}{2\pi}\right) w_{q\bar{q}}^1\,w_{q\bar{q}}^2\,w_{q\bar{q}}^3\,\widetilde{\Xi}_3^{\mathrm{CL,k}_t}(R)\,.
\end{align}
\end{subequations}

\begin{center}
\begin{figure}[ht]
\centering
\includegraphics[width=0.45\textwidth]{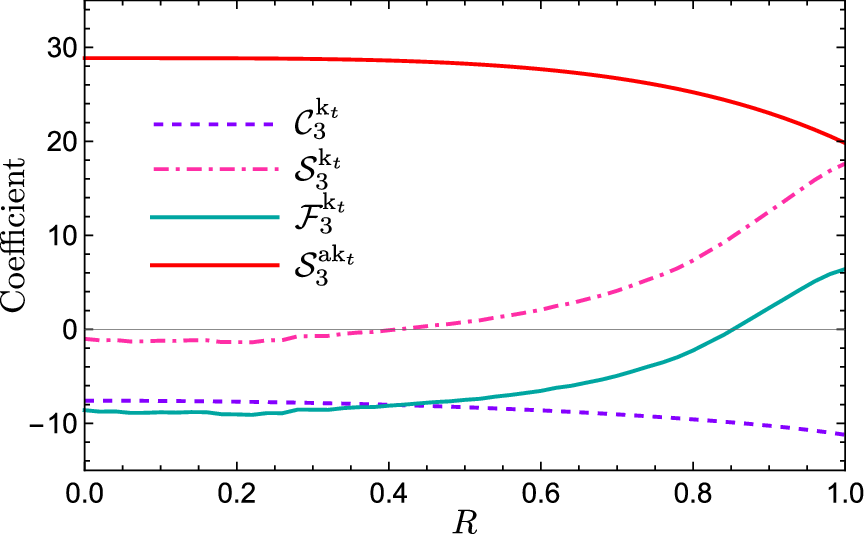}
\caption{\label{fig:02}CLs and NGLs coefficients at three loops with $k_t$ and anti-$k_t$ clustering.}
\end{figure}
\end{center}

We show in figure \ref{fig:02} a plot of the coefficients of NGLs and CLs in the $k_t$ and anti-$k_t$ algorithms as a function of the jet radius $R$. Shown also is the combined coefficient $\mathcal{F}^{\mathrm{k}_t}_3=\mathcal{S}_3^{\mathrm{k}_t}+ \mathcal{C}_3^{\mathrm{k}_t}$ for the $k_t$ algorithm.

As in the previous section, we notice that the NGLs coefficient in the anti-$k_t$ algorithm at this order is also quite large. Clearly, the application of the $k_t$ clustering has reduced the significance of NGLs by almost a factor of 30 for values less than $R\sim0.6$. The CLs coefficient is also small such that the overall coefficient $\mathcal{F}_3$ is smaller in magnitude than the $\mathcal{S}_3^{\mathrm{ak}_t}$ by about a factor of 3 for most values of $R$.

\section{Four loops and beyond}

\subsection{Four-loops NGLs with anti-\texorpdfstring{$k_t$}{kt} at small \texorpdfstring{$R$}{R}}

The calculation of NGLs with anti-$k_t$ clustering proceeds in a similar manner at fourth order, and can easily be deduced from previous calculations of NGLs in the literature. In fact, the phase space of integration is similar to that of the hemisphere mass distribution reported in ref. \cite{Khelifa-Kerfa:2015mma}, and thus the cumulative cross-section, at this order, may be cast in the following way
\begin{equation}
\Sigma_4^{\mathrm{ak}_t}(\Delta) = \frac{1}{4!}\left[\Sigma_1(\Delta)\right]^4+\frac{1}{2!}\left[\Sigma_1(\Delta)\right]^2\Sigma_2^{\mathrm{NG,ak}_t}(\Delta) +\Sigma_1(\Delta)\,\Sigma^{\mathrm{NG,ak}_t}_3(\Delta)+\frac{1}{2!}\left[\Sigma_2^{\mathrm{NG,ak}_t}(\Delta)\right]^2+\Sigma^{\mathrm{NG,ak}_t}_4(\Delta)\,,
\end{equation}
with pure NGLs contribution given by
\begin{subequations}
\begin{align}
\Sigma_4^{\mathrm{NG,ak}_t}(\Delta)&=\mathcal{S}_4^{\mathrm{ak}_t}(R)\,\frac{t^4}{4!}\,,\\
\mathcal{S}_4^{\mathrm{ak}_t}(R)&=2\,\CF\,\mathrm{C_A^3}\int\left(\prod_{i=1}^4\frac{\d c_i}{1-c_i^2}\,\frac{\d\phi_i}{2\pi}\right)
\Big[-\mathcal{A}_{q\bar{q}}^{12}\,\bar{\mathcal{A}}_{q\bar{q}}^{13}\,\bar{\mathcal{A}}_{q\bar{q}}^{14}\,\Theta_\inn(k_1)\Theta_\out(k_2)\Theta_\out(k_3)\Theta_\out(k_4)+\notag\\
&+3\,\bar{\mathcal{A}}_{q\bar{q}}^{12}\,\mathcal{B}_{q\bar{q}}^{134}\,\Theta_\inn(k_1)\Theta_\out(k_2)\Theta_\inn(k_3)\Theta_\out(k_4)
+\mathfrak{U}_{q\bar{q}}^{1234}\,\Theta_\inn(k_1)\Theta_\inn(k_2)\Theta_\out(k_3)\Theta_\out(k_4)-\notag\\
&-\mathcal{C}_{q\bar{q}}^{1234}\,\Theta_\inn(k_1)\Theta_\inn(k_2)\Theta_\inn(k_3)\Theta_\out(k_4)\Big]-\notag\\
&-2\,\CF\,\mathrm{C_A^3}\left(1-\frac{2\,\CF}{\CA}\right)\int\left(\prod_{i=1}^4\frac{\d c_i}{1-c_i^2}\,\frac{\d\phi_i}{2\pi}\right)\mathbf{A}_{q\bar{q}}^{1234}
 \Theta_\inn(k_1)\Theta_\inn(k_2)\Theta_\out(k_3)\Theta_\out(k_4)\,,\label{eq:NGLs4}
\end{align}
\end{subequations}
where the irreducible antenna functions read \cite{Delenda:2015tbo}
\begin{subequations}
\begin{align}
\mathcal{C}_{q\bar{q}}^{1234} & = w_{q\bar{q}}^1 \left(\mathcal{B}_{q1}^{234}+\mathcal{B}_{1\bar{q}}^{234}-\mathcal{B}_{q\bar{q}}^{234}\right),\\
\mathfrak{U}_{q\bar{q}}^{1234} & =w_{q\bar{q}}^1\left(\mathcal{A}_{q1}^{23}\,\bar{\mathcal{A}}_{q1}^{24} +\mathcal{A}_{1\bar{q}}^{23}\,\bar{\mathcal{A}}_{1\bar{q}}^{24} -\mathcal{A}_{q\bar{q}}^{23}\,\bar{\mathcal{A}}_{q\bar{q}}^{24}\right), \\
\mathbf{A}_{q\bar{q}}^{1234}&=
 \omega_{q\bar{q}}^1\,\omega_{q1}^2\left(\bar{\mathcal{A}}_{q\bar{q}}^{23}-\bar{\mathcal{A}}_{q1}^{23}\right)\left(\bar{\mathcal{A}}_{q1}^{24}-\bar{\mathcal{A}}_{1\bar{q}}^{24}\right)
+\omega_{q\bar{q}}^1\,\omega_{1\bar{q}}^2\left(\bar{\mathcal{A}}_{q\bar{q}}^{23}-\bar{\mathcal{A}}_{1\bar{q}}^{23}\right)\left(\bar{\mathcal{A}}_{1\bar{q}}^{24}-\bar{\mathcal{A}}_{q1}^{24}\right)-\notag\\
&-\omega_{q\bar{q}}^1\,\omega_{q\bar{q}}^2\left(\bar{\mathcal{A}}_{q1}^{23}-\bar{\mathcal{A}}_{q\bar{q}}^{23}\right)\left(\bar{\mathcal{A}}_{1\bar{q}}^{24}-\bar{\mathcal{A}}_{q\bar{q}}^{24}\right)
+k_3\leftrightarrow k_4 \,.
\end{align}
\end{subequations}

At small values of $R$ the integration has been performed in ref. \cite{Khelifa-Kerfa:2015mma}, and the corresponding result is given by
\begin{equation}
\mathcal{S}_4^{\mathrm{ak}_t}(R\sim0)=-\CF\,\mathrm{C_A^3}\,\zeta_4\left[\frac{29}{4}-\left(1-\frac{2\,\CF}{\CA}\right)\right].
\end{equation}

\subsection{LL resummation}

\begin{center}
\begin{figure}[ht]
\centering
\includegraphics[width=0.45\textwidth]{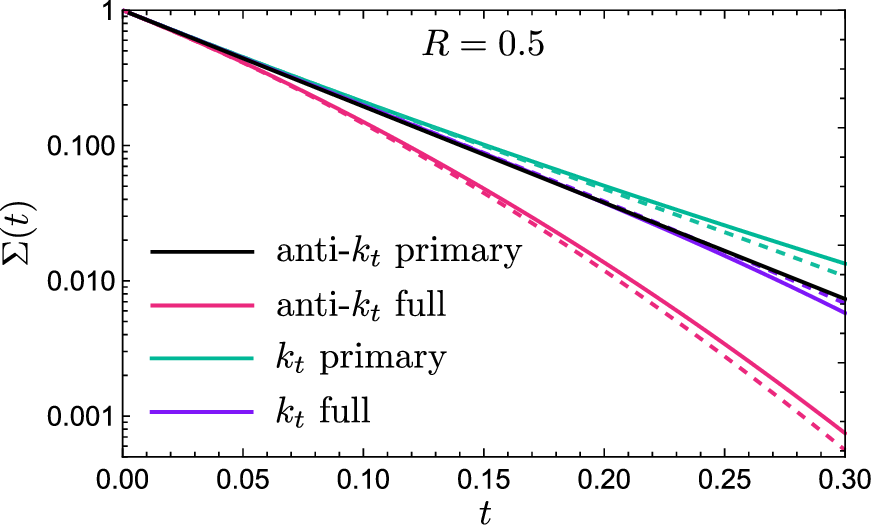}
\caption{\label{fig:NGLOR2}Numerically resummed NGLs and CLs at large $\Nc$.}
\end{figure}
\end{center}

In this section we present numerical results for the resummation of NGLs and CLs in the large-$\Nc$ approximation. For this we use the MC code first developed in refs. \cite{Dasgupta:2001sh,Dasgupta:2002bw} with modification of $k_t$ clustering in terms of distances \eqref{eq:dis}. The results are shown in figure \ref{fig:NGLOR2} for the resummed cumulative distribution $\Sigma$ as a function of the evolution parameter $t$ \eqref{eq:EvolParam}, for the particular value of the jet radius $R = 0.5$. Shown in the figure are: results for the (primary) global distribution (black curve), obtained by running the MC program using anti-$k_t$ clustering and allowing only for primary emissions, and the full anti-$k_t$ distribution (solid pink curve) which additionally includes NGLs at large $\Nc$. We observe the very large impact of NGLs on the distribution, reducing the global form factor by a factor of 10 for $t=0.3$.

We also show in the same figure the primary-emission distribution obtained by running the above-mentioned program with $k_t$ clustering (solid green curve), which includes the global form factor together with the resummed CLs, as well as the overall distribution in $k_t$ clustering which includes in addition the resummed NGLs (solid purple curve). We observe that the distribution in $k_t$ clustering is affected by both CLs and NGLs, but the combined impact of the two is noticeably small. This means that CLs tend to cancel NGLs in $k_t$ clustering at all orders, as noted with the fixed-order calculations performed above.

Moreover, we show in figure \ref{fig:NGLOR2} the analytical results for the resummed distribution in each case (dashed lines), which are estimated from the observed pattern of exponentiation
\begin{equation}
\Sigma(\Delta)=\exp\left[\Sigma_1(\Delta)\right]\exp\left[\sum_{i=2}^\infty \Sigma^{\mathrm{NG}}_i(\Delta)\right]\exp\left[\sum_{i=2}^\infty\Sigma^{\mathrm{CL}}_i(\Delta)\right].
\end{equation}
It is clear that the truncation of the series at $i=3$ in the exponent, though quite close to the numerical result, does not give an accurate fit of the MC distribution. This means that higher-order contributions cannot be ignored.

It is worth mentioning that the finite-$\Nc$ corrections to this distribution are negligible at LL accuracy, and the large-$\Nc$ limit should be a valid approximation for the purpose of comparison to experimental data. This was demonstrated in ref. \cite{Hatta:2013iba} for the gap-energy observable and in ref. \cite{Hagiwara:2015bia} for the jet mass distribution in $e^+e^-$ annihilation to dijets. However, finite-$\Nc$ corrections in non-global logarithms are expected to be important for observables measured at hadron colliders  \cite{Hatta:2020wre}.

\subsection{NLL resummation with anti-\texorpdfstring{$k_t$}{kt}}

We present here a resummed result for the azimuthal decorrelation distribution with anti-$k_t$ clustering at NLL accuracy in the large-$\Nc$ limit, obtained using the recently-published program \texttt{Gnole} \cite{Banfi:2021owj,Banfi:2021xzn}. The distribution can be obtained from this program by defining our observable within the code, using the full definition rather than the LL approximation \eqref{eq:def}. Explicitly written the former reads
\begin{equation}\label{eq:new}
\delta\phi = \left|\sin^{-1}\sum_{i\notin \mathrm{jets}}\frac{k_{\perp i}}{p_{tr}}\right|,
\end{equation}
where $k_{\perp i}$ is the component of the transverse momentum of the emission $i$ perpendicular to the thrust (or leading jet) axis, and $p_{tr}$ is the recoiling jet's transverse momentum. For simplicity we take the jets to be at threshold, i.e., transverse to the beam direction.
\begin{center}
\begin{figure}[ht]
\centering
\includegraphics[width=0.45\textwidth]{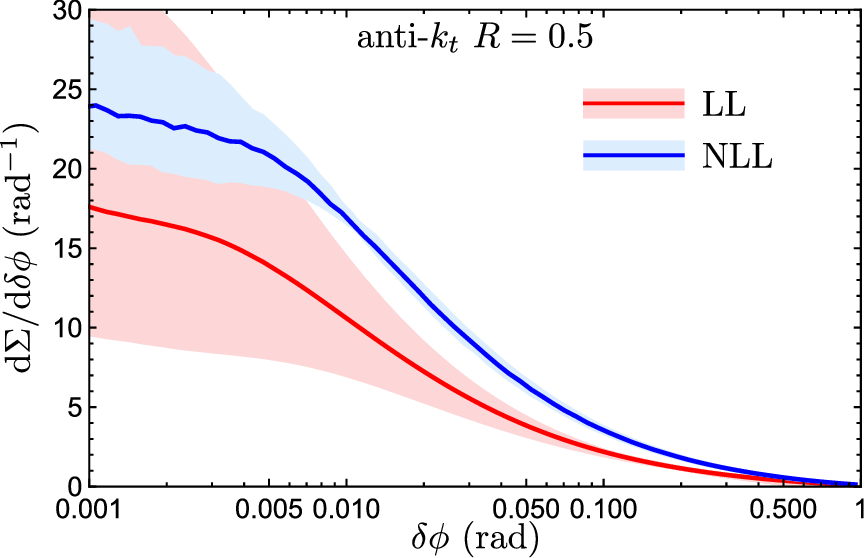}
\caption{\label{fig:NGLOR3} NLL numerical resummation of the $\delta\phi$ distribution at large $\Nc$.}
\end{figure}
\end{center}

In figure \ref{fig:NGLOR3} we present the results for the NLL resummed distribution together with the LL one, both obtained with \texttt{Gnole}. We note here that the LL distribution is obtained using the definition of the observable \eqref{eq:new}, while that obtained with the MC code of refs. \cite{Dasgupta:2001sh,Dasgupta:2002bw} (figure \ref{fig:NGLOR2}) is essentially equivalent to the transverse energy distribution (in other words, definition \ref{eq:def} without the $\sin\phi$ part). This does in fact numerically affect the distribution even at LL accuracy. \footnote{Note that it is not possible to change the definition of the observable in the MC code of refs. \cite{Dasgupta:2001sh,Dasgupta:2002bw}.}

We observe that the NLL corrections to the distribution are quite important, just like the transverse energy distribution shown in ref. \cite{Banfi:2021xzn}. Furthermore, the scale-uncertainty band, obtained by varying the renormalization and resummation scales by factors of $2$ and $1/2$ around their central values ($\mu_R=\sqrt{s}$ and $Q_0=\sqrt{s}/2$, respectively, with $\sqrt{s}=M_Z$), gets significantly reduced in the NLL curve.

We note that the actual distribution does not possess a Sudakov peak at low values of $\delta\phi$, instead it tends towards a constant value. This is explained by the fact that the very low values of $\delta\phi$ are not suppressed by soft emissions, but are rather enhanced by vectorial cancellation of semi-hard emissions.

It is worth mentioning that one still needs to account for the matching in order to fully control all sub-leading NLL logarithms at the tail of the distribution \cite{Banfi:2021owj,Banfi:2021xzn}. The procedure of matching ensures adding terms that are unaccounted for by the resummation, preferably up to NLO or even NNLO, and removing double-counted logarithmic terms already present in the resummation, in a way that the expansion of the matched distribution correctly reproduces the full NLO (or NNLO) result, while still controlling the NLL logarithms at all orders. This way one extends the validity of the resummed distribution at hand over the entire range of values of $\delta\phi$. Given the small scale uncertainties in the NLL resummation, this distribution (after matching) will be very valuable for collider phenomenology.

\section{Conclusions}

In this letter we have presented both fixed-order and all-orders results for the distribution of the azimuthal decorrelation observable for the specific QCD process $e^+ e^-$ annihilation into two jets. The said observable is of non-global nature and hence its distribution contains, in addition to the usual global logarithms, non-global and/or clustering logarithms. These logarithms are jet-algorithm dependent, and start to appear at two loops and are quite delicate to compute.

For NGLs, we have calculated at finite $\Nc$ the full-$R$ expression analytically at two loops and numerically at three loops for the anti-$k_t$ jet algorithm. At four loops, they have been determined only for small-$R$ values for the same said jet algorithm. For the $k_t$ clustering algorithm, calculations have been performed up to three loops and only numerically.

Moreover, CLs, which are absent in the anti-$k_t$ algorithm, have been computed numerically up to three loops. The usual reduction in the significance of NGLs due to $k_t$ clustering has been observed confirming previous findings. Furthermore, the combined impact of NGLs and CLs on the distribution is observed to be very small which has important phenomenological implications in terms of accuracy of the resummed distribution. Our results at two loops have been checked against the output of the MC program \texttt{EVENT2}.

Numerical estimates of the all-orders distribution in the large-$\Nc$ approximation have been presented both at LL and NLL accuracy. The achievement of the latter accuracy has been made possible by the recently-published \texttt{Gnole} code. NLL resummation exhibits a better distribution both in terms of accuracy and scale uncertainty. It is thus worth investigating the impact of NLL effects with $k_t$ clustering. In our estimation, since the size of NGLs is significantly reduced with $k_t$ clustering, we expect that all sources of error in the distribution, including NLL corrections, finite-$\Nc$ corrections, and scale uncertainties, will not be as large as those in the anti-$k_t$ algorithm. This can be confirmed once a modified version of \texttt{Gnole} with $k_t$ clustering is made available. Of similar worthiness is computing NGLs and CLs at four loops with full-$R$ dependence.

\section*{Acknowledgements}

This work is supported by PRFU research project B00L02UN050120230003. The authors wish to thank the Algerian Ministry of Higher Education and Scientific Research and DGRSDT for
financial support.

The numerical calculations presented here have been performed in the HPC cluster at the University of Batna 2 (UB2-HPC).

We thank Andrea Banfi for clarifications about using \texttt{Gnole} program.

\bibliographystyle{model1a-num-names}

\bibliography{mybibfile}

\end{document}